\documentclass[universe,article,accept,pdftex,oneauthor]{Definitions/mdpi} 

\firstpage{1} 
\pubvolume{0}
\issuenum{0}
\articlenumber{0}
\pubyear{2024}
\copyrightyear{2024}
\externaleditor{Academic Editor: Firstname\linebreak Lastname}
\datereceived{21 March 2024} 
\daterevised{11 April 2024} 
\dateaccepted{13 April 2024} 
\datepublished{ }

\hreflink{https://doi.org/} 

\usepackage{bm}
\usepackage{amsmath,amssymb}
\usepackage{wasysym}
\usepackage{pifont,bbm}
\usepackage{stmaryrd}
\newcommand{\xxx}{y_1}
\newcommand{\yyy}{y_2} 
\newcommand{\Ua}{U_{{\rm A} a}}
\newcommand{\apj}{Astroph. J.}

\newcommand{\mnras}{Mon. Not. R. Astron. Soc.}
\newcommand{\aap}{Astron. Astrophys.}
\newcommand{\be}{\begin{equation}}
\newcommand{\ee}{\end{equation}}

\makeatletter
\renewcommand{\maketag@@@}[1]{\hbox{\m@th\normalsize\normalfont#1}}
\makeatother

\Title{Linear Stability Analysis of Relativistic Magnetized\linebreak Jets: {The Minimalist Approach}}

\TitleCitation{Linear Stability Analysis of Relativistic Magnetized Jets: The Minimalist Approach}

\Author{Nektarios Vlahakis \orcidA{}}

\AuthorNames{Nektarios Vlahakis}

\AuthorCitation{Vlahakis, N.}

\address[1]{%
Department of Physics, National and Kapodistrian University of Athens, University Campus,\linebreak Zografos, GR-157 84 Athens, Greece; vlahakis@phys.uoa.gr 
}

\abstract{A minimalist approach to the linear stability problem in fluid dynamics is developed that ensures efficiency by utilizing only the essential elements required to find the eigenvalues for given boundary conditions. It is shown that the problem is equivalent to a single first-order ordinary differential equation, and that studying the argument of the unknown complex function in the eigenvalue space is sufficient to find the dispersion relation. The method is applied to a model for relativistic magnetized astrophysical jets.
}


\begin{document}

\section{Introduction}\label{introduction}

Understanding the stability properties of magnetized plasma flows is important for unraveling the basic characteristics of many phenomena we observe in astrophysics, solar, and space physics. Unstable perturbations may significantly alter the dynamics of these flows and be the reason behind changes in the shape; bulk velocity; magnetization; heating; particle acceleration; and consequently, the emitted radiation.

Linear stability analysis is a tool that enables simplifying the problem as much as possible, focuses on the main ingredients of each mechanism, and corroborates more complicated and expensive work, such as numerical simulations or laboratory experiments.    
It has its own difficulty, mainly due to the complex mathematics involved, especially in the relativistic regime, and using curvilinear coordinates. 
This remains true even if one neglects dissipative effects related to finite viscosity and resistivity, and
assumes a simplified unperturbed state of a cylindrical flow with helical magnetic field, which approximates astrophysical jets in the propagation phase well. 

From the simplest possible cases we know of, there exist unstable modes related to discontinuities of density (Rayleigh--Taylor), shear of bulk velocity (Kelvin--Helmholtz), current (current driven), or rotation (centrifugal and magnetorotational). An overview of the linear stability method and many nonrelativistic cases can be found in Refs.~\cite{1961hhs..book.....C,Goedbook2}.
In magnetized astrophysical jets, the various instabilities often appear simultaneously, and it is nontrivial to disentangle them. 
Studies for nonrelativistic magnetized jets can be found, e.g., in Refs. \cite{FTZ81,1983ApJ...269..500C,AC92,ALB00,2011A&A...525A.100B,2013MNRAS.434.3030B,2016MNRAS.462.3031B}. There is also a relatively small number of studies that examined relativistic magnetized jets \cite{Hardee07,2013MNRAS.434.3030B,2017MNRAS.467.4647K,2018MNRAS.474.3954K,2019MNRAS.485.2909B} and two more recent studies \cite{2023MNRAS.523.6294S,2023A&A...680A..46S} based on the methodology of the full problem presented in \cite{2023Univ....9..386V}.
Refs. \cite{IP96,2009ApJ...697.1681N,2017MNRAS.468.4635S,2019MNRAS.482.2107D} also consider relativistic magnetized jets in the force-free limit.  

All existing works on ideal fluids (neglecting dissipative effects) share a common procedure: they arrive in a system of two first-order differential equations in the complex domain, or equivalently, one second-order differential equation. The requirement that the solution of this system needs to satisfy certain boundary conditions gives the dispersion relation. 
In this paper, we show that this procedure may be simplified by reducing the number of equations by half. Following this novel approach, which we dub minimalist, one needs to solve only one differential equation in order to find the eigenvalues of the problem. 
In Section~\ref{secminimalist}, we present the equations of the approach, and in Section~\ref{seceigenarg}, we explain how to solve the boundary condition, again following a minimal path by using only the argument of the eigenfunction. 
In Section \ref{examplecase}, we apply the method to a model of a magnetized jet, extending a result of \cite{ALB00} to relativity. We conclude with a discussion in Section \ref{secconclusions}.

\section{The Minimalist Approach}\label{secminimalist}

The linearization of the relativistic ideal magnetohydrodynamic equations in cylindrical coordinates $(\varpi,\phi,z)$, assuming a cylindrical unperturbed state and perturbations of the form $Q_1(\varpi) \exp\left[i\left(m\phi +kz-\omega t\right)\right]$ for all physical quantities, yields the following system of two complex first-order differential equations: 
\begin{eqnarray}
	\dfrac{d}{d\varpi} \left( \begin{array}{c}
		\xxx \\ \yyy
	\end{array} \right) + \dfrac{1}{{\cal D}} \left( \begin{array}{cc}
		{\cal F}_{11} & {\cal F}_{12} \\ {\cal F}_{21} & {\cal F}_{22}
	\end{array} \right) \left( \begin{array}{c}
		\xxx \\ \yyy
	\end{array} \right)
	=0\,.
	\label{systemodes}
\end{eqnarray}

Here $\xxx$ is related to the Lagrangian displacement in the radial direction, and $\yyy$ is related to the total pressure in the perturbed locations of fluid elements. There are known algebraic relations giving all the other quantities in terms of $\xxx$ and $\yyy$, and thus, these two functions fully determine the solution. 
The pair of functions $\xxx$ and $\yyy$ is the most convenient choice because they are continuous everywhere, even at cylindrical surfaces, where the undisturbed state is discontinuous, e.g., in the interface between the jet and its environment. 
Details on the form of the unperturbed state, the derivation of the system of Equation~(\ref{systemodes}), the expressions for the various ${\cal F}_{ij}/{\cal D}$, and the proof that $\xxx$ and $\yyy$ should be continuous everywhere can be found in \cite{2023Univ....9..386V}.

The above system is linear, and if our purpose is to find the eigenvalues and eigenfunctions, the proportionality constants are free. The only constraints that need to be satisfied are the boundary conditions. All kinds of boundary conditions are discussed in \cite{2023Univ....9..386V} and they refer to either on the axis, an interface where the unperturbed state is discontinuous, infinity, or a solid boundary (wall). In all these cases, the boundary conditions give the~ratio 
\be Y=\dfrac{\xxx}{\yyy}
\label{eqdefY}
\ee 
and not the functions $\xxx$ and $\yyy$ separately. 
Thus, a minimalist approach is to work with this ratio, which has the advantage that it is uniquely defined for each eigenvalue, but most importantly, that the number of differential equations is reduced by half. The new equation is non-linear, but this does not burden the procedure if one satisfies the boundary conditions using a shooting method. 

\subsection{Differential Equation}

The differential equation for the new unknown function $Y$ can be derived by direct differentiation and using the system~(\ref{systemodes}). It is the following single complex equation: 
\begin{eqnarray}
	\dfrac{dY}{d\varpi}=\dfrac{{\cal F}_{21}}{{\cal D}}Y^2+\dfrac{{\cal F}_{22}-{\cal F}_{11}}{{\cal D}}Y-\dfrac{{\cal F}_{12}}{{\cal D}}
	\,. \label{eqforY} \end{eqnarray}

Any solution of this equation that satisfies certain boundary conditions, which will be discussed below, and is continuous everywhere is an eigenfunction corresponding to a particular eigenvalue.
The original pair of functions, if needed, can be found a posteriori using the equations
(that are equivalent to the original system)
\be 
\dfrac{\yyy'}{\yyy}=-\dfrac{{\cal F}_{21}}{{\cal D}}Y-\dfrac{{\cal F}_{22}}{\cal D}  
\,, \quad \dfrac{ \xxx'}{\xxx}=-\dfrac{ {\cal F}_{12}}{{\cal D}}\dfrac{1}{Y}-\dfrac{ {\cal F}_{11}}{\cal D} \,.
\label{eqforxxxyyy}
\ee

(Note that only one variable, i.e., $\yyy$, needs to be found by solving the related differential equation, the other is given by $\xxx=Y\yyy$.)

An instructive approximate solution of Equation (\ref{eqforY}), which shows its typical expected behavior, is presented in Appendix \ref{appendixapprox}.

\subsection{Boundary Conditions on the Axis}

{On the axis, there are regularity conditions on $\xxx$ and $\yyy$ that are translated in a condition on their ratio $Y$. 
The detailed derivation of the former can be found in \cite{2023Univ....9..386V}. In addition, the behavior of $Y$ near the axis is discussed in Appendix \ref{secboundaxis}.

The resulting boundary condition at the symmetry axis is 
\be 
\mbox{for } m\neq 0\,, \quad 
Y(\varpi=0)=\displaystyle\lim_{\varpi\to 0}\dfrac{\varpi {\cal F}_{11}-|m|}{\varpi {\cal F}_{21}}=-\displaystyle\lim_{\varpi\to 0}\dfrac{\varpi {\cal F}_{12}}{\varpi {\cal F}_{11}+|m|}=\dfrac{\lambda_1}{\lambda_2}
\,, \label{eqbcaxismneq0} \ee 
\be \mbox{for } m=0\,, \quad Y(\varpi=0)= -\displaystyle\lim_{\varpi\to 0}\dfrac{\varpi {\cal F}_{12}}{2}=-\dfrac{b_{12}}{2}\varpi^2 \approx 0
\,. \label{eqbcaxismeq0} \ee 
}

\subsection{Boundary Conditions at Infinity}

{If at large distances from the axis, the medium is static and homogeneous with zero $B_{0\phi}$, as it is usually assumed, the solution for $Y$ can be found analytically; see  
Appendix~\ref{secboundinfinity}. 
In any case, whatever the unperturbed state of the jet environment, care must be taken that the perturbations vanish as $\varpi\to\infty$, and if they are oscillating, they correspond to outgoing waves in the radial direction. 
}

\subsection{Boundary Conditions at Interfaces}

The function $Y$ is everywhere continuous, including possible interfaces where the unperturbed state is discontinuous. 

In the subcase of a solid boundary (wall), we simply require $Y=0$ at the boundary. 

\section{Finding the Eigenvalues}\label{seceigenarg}

The dispersion relation is found by solving Equation (\ref{eqforY}) subject to boundary conditions. 
In the following, we use the temporal approach (given a real $k$ and integer $m$, search for complex eigenvalues $\omega$), although the procedure is the same in the spatial approach (given a real $\omega$ and integer $m$, search for complex eigenvalues $k$).

Also, to be specific, suppose we have a jet with radius $\varpi_j$ and we know $Y$ in its environment as a function of $\varpi$, $k$, $m$, and $\omega$. Its value $Y|_{\varpi=\varpi_j^+}$ on the boundary is denoted as $Y_{\rm BC}$ and is an analytic function of the complex variable $\omega$.
For given $m$, $k$, and all possible $\omega$, we can integrate in the interior of the jet, starting from the axis using the appropriate boundary condition there, and when we reach the interface $\varpi=\varpi_j$ from the left, we find the value $Y|_{\varpi=\varpi_j^-}$, which, for brevity in this section, we simply call $Y$. This is also an analytical function of the complex variable $\omega$. The accepted eigenvalues $\omega$ are the ones for which $Y=Y_{\rm BC}$. 

This procedure with obvious modifications can be applied to any other case. For example, if we have discontinuities in the jet interior, we simply cross them, keeping $Y$ continuous and again work with $Y=Y_{\rm BC}$ at the jet surface. 
If we do not know the solution in the environment, we simply continue the integration in the $\varpi>\varpi_j$ regime, and the boundary condition that will determine the dispersion relation is at infinity. Alternatively, we begin the integration from a large distance, reach $\varpi_j$ from the right, and match with the solution from the left. 
In any case, a complex equation of the form $Y=Y_{\rm BC}$ at some surface will determine the eigenvalues.

\subsection*{Roots and Poles as Positive and Negative Line Charges}

It is essential to work with the difference $Y-Y_{\rm BC}$ in the complex plane $\omega$. Since we are interested in unstable modes, we consider only the half plane $\Im\omega>0$. The properties of this analytic function  help us to find the eigenvalues, which are its roots.  
It is convenient to write it as $Y-Y_{\rm BC}=e^{-\Phi+i\Psi}$, where $-\Phi$ and $\Psi$ are the real and imaginary parts of $\ln(Y-Y_{\rm BC})$. Equivalently, $\Phi=-\ln |Y-Y_{\rm BC}|$ and $\Psi={\rm Arg}[Y-Y_{\rm BC}]$.

There is a direct analogy between the complex plane $\omega$ and a Cartesian $xy$ plane by writing $\omega=x+iy$. 
The Cauchy--Riemann conditions give $\nabla \Phi\bot \nabla \Psi$ and that $\Phi$ and $\Psi$ satisfy the Laplace equation at all points $y>0$, except the positions where $\Phi$ becomes infinity, i.e., the roots and poles of the function $Y-Y_{\rm BC}$.

Suppose this function has roots $\omega_{rn}$, $n=1,2,\dots$, and poles 
$\omega_{pm}$, $m=1,2,\dots$,
with the corresponding positions in the $xy$ plane $\bm r_{rn}$ and $\bm r_{pm}$. 
The function $\Phi(x,y)$ can be thought of as an electric potential associated with line charges, i.e., positive at the positions of the roots and negative at the positions of the poles (sources of the potential are also line charges located at positions with $\Im\omega\le 0$, but we are interested in finding the line charges only at $\Im\omega>0$). The function $\Psi$ is the stream function of the corresponding electric field $-\nabla \Phi=\nabla\Psi\times\hat z$ (the field lines are isocontours of $\Psi$ and are normal to the isopotentials $\Phi=$ constant).\endnote{{There are other ways to make connections with other physical settings. We can think of the roots as line sources of incompressible fluid and the poles as line sinks. In another analogy, we can think of the roots/poles as line vortices with positive/negative circulation, respectively.
Another possibility is to treat the real/imaginary parts of $Y-Y_{\rm BC}$ as a potential/stream function. In this picture, there is an electric cylindrical dipole at the location of each pole and the potential/stream function vanishes at the positions of the roots.
}} 

These can be proved by looking at the form of the potential/stream function near roots and poles.
Writing the function as $Y-Y_{\rm BC}=C_{rn}(\omega) (\omega-\omega_{rn})^{q_{rn}}$
(with $q_{rn}$ as the multiplicity of the root),
we indeed see that the dominant contributions near a root are
$\Phi\approx -q_{rn}\ln |\bm r-\bm r_{rn}|+C_{rn}$ and $\Psi\approx q_{rn} \arctan\dfrac{y-y_{rn}}{x-x_{rn}}+D_{rn}$ (where $C_{rn}$ and $D_{rn}$ are constants),
corresponding to an electric field of a positive line charge $-\nabla \Phi=\nabla\Psi\times\hat z 
\approx q_{rn}\dfrac{\bm r-\bm r_{rn}}{|\bm r-\bm r_{rn}|^2}$.
Similarly, near a pole, by writing the function as $Y-Y_{\rm BC}=\dfrac{C_{pm}(\omega)}{ (\omega-\omega_{pm})^{q_{pm}}}$, 
we find that the dominant contributions near the pole are $\Phi\approx +q_{pm}\ln |\bm r-\bm r_{pm}|+C_{pm}$ and $\Psi\approx -q_{pm} \arctan\dfrac{y-y_{pm}}{x-x_{pm}}+D_{pm}$ (where $C_{pm}$ and $D_{pm}$ are constants),
corresponding to an electric field of a negative line charge $-\nabla \Phi=\nabla\Psi\times\hat z 
\approx -q_{pm}\dfrac{\bm r-\bm r_{pm}}{|\bm r-\bm r_{pm}|^2}$.

Near each positive line charge (a root), the relations 
$\Phi\approx -q_{rn}\ln |\bm r-\bm r_{rn}|+C_{rn}$ and $\Psi\approx q_{rn}\arctan\dfrac{y-y_{rn}}{x-x_{rn}}+D_{rn}$ mean that the polar coordinates in a system with the axis at the line charge are $ e^{-\Phi+C_{rn}}$ and $\Psi-D_{rn}$. It is important to note, first, that $\Phi$ is $+\infty$ at the location of the line charge and decreases as we move away, and second, that $\Psi$ increases as we move counterclockwise around the line charge.
Working similarly, near a negative line charge (a pole), we conclude that we have the opposite behavior: $\Phi$ is $-\infty$ at the location of the line charge and increases as we move away, and $\Psi$ decreases as we move counterclockwise around the line charge. 

Evidently, the potential and the stream function are not independent. We can find one from the other using the Cauchy--Riemann conditions and their isocontours are normal to each other. In the spirit of the minimalist approach, we can use only one. The stream function is a better choice since it does not involve infinities.

\begin{figure}[H]
	\begin{minipage}{0.63\textwidth} 
		\includegraphics[width=.85\textwidth]{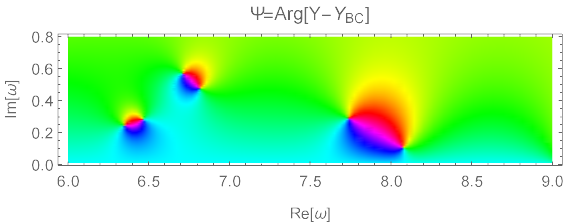}\hspace{-6pt}
		
		\vspace{3mm} 
		\includegraphics[width=.85\textwidth]{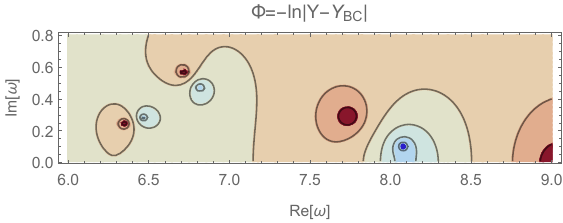} \hspace{-6pt}
	\end{minipage} 
	\hfill
	\begin{minipage}{0.33\textwidth} 
		
		\vspace{6mm}
		\hfill
		\includegraphics[width=1\textwidth]{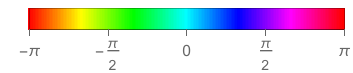}
		
		\vspace{28mm}
		\hfill 
		\includegraphics[width=1\textwidth]{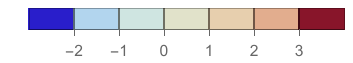}
		
		\vspace{4mm} $\ $
	\end{minipage}
	
	\includegraphics[width=0.29\textwidth]{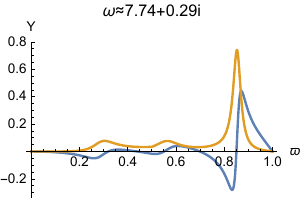}\includegraphics[width=0.29\textwidth]{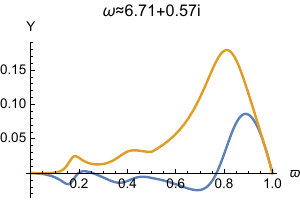}\includegraphics[width=0.41\textwidth]{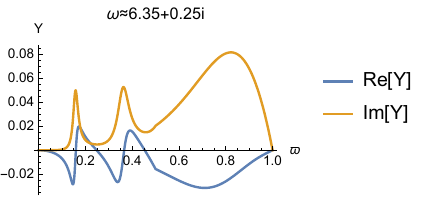}
	\caption{The two upper panels show the parts of the function $Y-Y_{\rm BC}$ in the complex $\omega$ plane, corresponding to the case examined in \cite{2023Univ....9..386V}. $Y$ is the value at the jet radius $\varpi=1$ as the result of the integration from the axis, and $Y_{\rm BC}=0$ in this particular case.
	The bottom row shows the eigenfunctions for the three eigenvalues that satisfy $Y-Y_{\rm BC}=0$.} 
	\label{figwebpapertheory}
\end{figure}

Concluding, and returning to the $\omega$ plane, in order to find the eigenvalues, we only need the isocontours of $\Psi={\rm Arg}[Y-Y_{\rm BC}]$. 
An example is shown in the upper panel of Figure~\ref{figwebpapertheory}.
The points where $\Psi$ experiences jumps equal to $\pi$ are either roots or poles. 
Moving counterclockwise around such a point, if $\Psi$ increases, it is a root, and if $\Psi$ decreases, it is a pole. 
Moving from right to left, we see six line charges at locations
$\omega\approx 8.1+0.1i$,
$\omega\approx 7.74+0.29i$,
$\omega\approx 6.82+0.47i$,
$\omega\approx 6.71+0.57i$, 
$\omega\approx 6.46+0.28i$, and
$\omega\approx 6.35+0.25i$. 
Checking the change in $\Psi$, we conclude that the first is a pole, the second a root, etc.

\begin{figure}[H]
	\begin{minipage}{0.58\textwidth}
		\includegraphics[width=.9\textwidth]{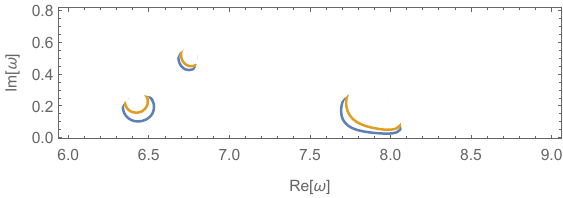}
		
		\vspace{3mm} 
		\includegraphics[width=.9\textwidth]{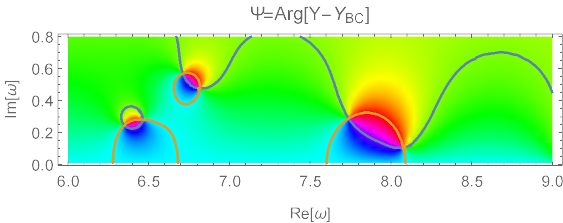}
		
		\vspace{3mm} 
		\includegraphics[width=.9\textwidth]{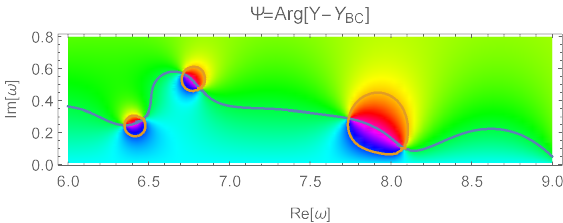} 
	\end{minipage} 
	\hfill
	\begin{minipage}{0.35\textwidth} 
		
		\vspace{46mm}
		\hfill
		\includegraphics[width=1\textwidth]{Figures/colorbar_arg}

		\vspace{4mm} $\ $
	\end{minipage}
	\caption{Two contours are shown in the {\bf upper panel}, i.e., $\Psi_1=0.3$ (blue) and $\Psi_2=0.6$ (orange).
	The Spectral Web in the {\bf middle panel} shows the contours where $Y-Y_{\rm BC}$ is real (with blue) and purely imaginary (with orange).
	The {\bf lower panel} shows (with blue) the field line that connects all the roots and poles, which corresponds to real $(Y-Y_{\rm BC})e^{i\pi/4}$. The purely imaginary $(Y-Y_{\rm BC})e^{i\pi/4}$ is also shown (in orange).} 
	\label{figwebpapertheorynew}
\end{figure}

In fact, even the full map of isocontours of $\Psi={\rm Arg}[Y-Y_{\rm BC}]$ is not necessary. If we are in a position at the $\omega$ plane, find its $\Psi$, and then find the $\nabla \Psi$ at this point (by looking the $\Psi$ of its neighbors), we know that a positive charge can be found in the direction opposite to the field $\nabla\Psi\times\hat z$.
We can move from point to point on the curve $\Psi=$ constant and we will reach a positive charge.
In another variant that is probably even more economic, we can plot the isocontours of just two neighboring values $-\pi<\Psi_1<\Psi_2\le \pi$. The roots/poles are the endpoints of the contours, while the direction of $\nabla\Psi$ (it is normal to the contours pointing from the $\Psi_1$ to the $\Psi_2>\Psi_1$ contour) is sufficient to understand which end is the root (it is in the direction opposite to the field $\nabla\Psi\times\hat z$). An example is shown in the upper panel of Figure~\ref{figwebpapertheorynew}.
\vspace{-10pt}
\vspace{-10pt}

The ``Spectral Web'' method was developed and presented in \cite{2018PhPl...25c2109G,Goedbook2} for nonrelativistic magnetohydrodynamic flows. According to this method, in cases where $y_{1\rm BC}=0$, the eigenvalues can be found 
in the $\omega$ complex plane
as intersections of the ``solution paths'', where $\left\{\Re [\xxx] \Im[\yyy] -\Im [\xxx] \Re[\yyy]\right\}_{\rm BC}=0$, and ``conjugate paths'', where $\left\{\Re [\xxx] \Re[\yyy]+\Im [\xxx] \Im[\yyy]\right\}_{\rm BC}=0$. 
This system gives the roots but also ``spurious roots'' corresponding to $y_{2\rm BC}=0$.
They also discuss the ``complex oscillation theorem'', according to which, along the solution path and in between the spurious roots, $\dfrac{\xxx}{\varpi\yyy}$, which they call the alternator, is real and monotonic function of the arc length, and similarly along the conjugate path, the alternator is purely imaginary and monotonic function of the arc~length.

The analogy with the minimalist approach that uses the analytical function $Y$ is obvious: the ``solution paths'' correspond to $\Im [Y-Y_{\rm BC}]=0\Leftrightarrow {\rm Arg} [Y-Y_{\rm BC}]=0$ or $\pi$, the ``conjugate paths'' correspond to $\Re [Y-Y_{\rm BC}]=0\Leftrightarrow {\rm Arg} [Y-Y_{\rm BC}]=\pm \pi/2$ (these paths are shown in the middle panel of Figure~\ref{figwebpapertheorynew}), the roots correspond to positive line charges, and the spurious roots to negative line charges (poles).
As we move along a field line $\Psi={\rm Arg} [Y-Y_{\rm BC}]=$ constant (any constant, not only $0$, $\pm \pi/2$, and $\pi$) approaching a positive line charge, the potential
$\Phi=-\ln|Y-Y_{\rm BC}|$ monotonically increases. At the position of the positive line charge, the potential becomes $+\infty$ and the argument $\Psi$ experiences a jump equal to $\pi$ (since it corresponds to the angular polar coordinate in the local system with the line charge at the axis). As we move away from the positive line charge, the potential decreases, reaching $-\infty$ if we meet a negative line charge.

As suggested by \cite{Goedbook2}, this property helps with counting the eigenmodes that correspond to roots we meet as we move along a single field line. 
However, there is not always a single line connecting all the roots, making the counting of solutions impossible in general. 
In fact, in the particular case, there is a single line connecting all roots and poles, but it is not the ``solution path'' nor the ``conjugate path''. It is the line in which $(Y-Y_{\rm BC})e^{i\pi/4}$ is real, consisting of the parts ${\rm Arg}[Y-Y_{\rm BC}]=-\pi/4$, $3\pi/4$ shown in the lower panel of Figure~\ref{figwebpapertheorynew} with blue (the normal field lines corresponding to ${\rm Arg}[Y-Y_{\rm BC}]=\pi/4$, $-3\pi/4$ are also shown with orange lines).

The ``Spectral Web'' corresponds to the field lines ${\rm Arg}[Y-Y_{\rm BC}]=0$, $\pi/2$, as well as their continuations after jumps ${\rm Arg}[Y-Y_{\rm BC}]=\pi$, $-\pi/2$. Although these values do not seem to have any special significance relative to others, it is convenient to include these ``paths'' in the $\omega$ plane to show the location of line charges through the crossings of the paths (as already discussed, any other choice of isocontours will also show these locations). We include them in the following and continue to name the plot a ``Spectral Web'', although it has important additional information. 
The paths themselves are not enough; we need to know the values of $\Psi$ and separate roots from poles knowing whether $\Psi$ increases or decreases when moving counterclockwise around a line charge (or equivalently, understand the direction of $\nabla\Psi$ and the electric field $\nabla\Psi\times\hat z$). 

The middle panel of Figure~\ref{figwebpapertheory} shows the potential for illustrative purposes (since the eigenvalues were already found from the contours of $\Psi$ alone). We verify that $\Phi=+\infty$ at the roots and $\Phi=-\infty$ at the poles. We also see that the isopotentials are normal to the field lines shown in the upper panel as contours of $\Psi$ (to see the angles correctly, it is necessary to have the same scaling in the $\Re\omega$ and $\Im\omega$ axes).
The bottom panels of Figure~\ref{figwebpapertheory} show the eigenfunctions for the three eigenvalues found.
All the results of that figure correspond to the case examined in Section 6 of \cite{2023Univ....9..386V}.

{A last point to discuss is related to the possibility to encounter infinities of $Y$ when we integrate Equation~(\ref{eqforY}). Since $Y=\xxx/\yyy$ is defined as a ratio, one could expect infinities at points where $\yyy=0$. However, working in the complex domain, an infinite value of $Y$ requires both the real and imaginary parts of $y_2$ to vanish simultaneously with perfect accuracy, which is something that never happens during a numerical integration. Even if we encounter such a point, we automatically pass through it without a problem. An example is shown in Appendix \ref{appendixpole}.}

\section{Energy Consideration}\label{secenergy}

In the minimalist approach, where we do not solve an equation reminiscent of Newton's second law, one may think that the connection with the energy principle that is often used to discuss instabilities (through the increase/decrease of kinetic energy due to a decrease/increase of potential when moving from an equilibrium, in analogy to a ball on a hill/valley; see, e.g., \cite{2005ppa..book.....K}) has been lost. However, there is still some connection, as one could expect. 

The energy of a particular mass of the plasma in general evolves in time according to the relation $\dfrac{d{\cal E}}{dt}=-\displaystyle\oiint \left[
\Pi\bm V-(\bm V\cdot\bm B)\bm B\right] \cdot d\bm a$, which can be easily proved using the equations of motion\endnote{{The proof can be performed by writing the equations of motion as $T^{\mu\nu}_{\ ;\nu}=0$ and elaborating the energy momentum tensor, whose components in Cartesian coordinates are $ T^{00}=\gamma^2 \xi \rho_0 -P + \displaystyle\frac{E^2+B^2}{2}$,
		$T^{0j}=T^{j0}=\left( \xi \rho_0 \gamma^2 {\bm V}+
		{\bm E} \times {\bm B}\right) \cdot \hat{x}_j$,
		$T^{ij}=\xi \rho_0 \gamma^2 {V_i V_j }-
		E_i E_j + B_i B_j +\left(P+ \displaystyle\frac{ E^2 + B^2}{2}
		\right) \delta^{ij} $, with $i, j=1,2,3$.
		The equation for the energy is 
		$\dfrac{\partial T^{00}}{\partial t}+\nabla \cdot \left(T^{0i}\hat x_i\right)=0$. Its integral form in a volume whose boundary is moving with velocity $\bm V_s$
		(and thus, each part of the boundary creates a volume $\bm V_s dt \cdot d\bm a$ in the time interval $dt$) is
		$\dfrac{d}{dt}\displaystyle\iiint T^{00}d\tau +\displaystyle\oiint \left(T^{j0}\hat x_j-T^{00}\bm V_s\right)\cdot d\bm a=0$.
		Following the volume of a given mass, each point of the boundary moves with $\bm V_s=\bm V$, and substituting the components of the tensor, we obtain $\dfrac{d{\cal E}}{dt} +\displaystyle\oiint \left( 
		{\bm E} \times {\bm B}+P\bm V-\displaystyle\frac{ E^2 + B^2}{2}\bm V\right)\cdot d\bm a=0$.
		Substituting $\bm E=-\bm V\times \bm B$, we arrive at 
		$\dfrac{d{\cal E}}{dt}=-\displaystyle\oiint \left[
		\Pi\bm V-(\bm V\cdot\bm B)\bm B\right]\cdot d\bm a$, with $\Pi=P+\dfrac{B^2-E^2}{2}$.
}}. 
Considering the volume of the jet and using the boundary condition on its perturbed surface, according to which the magnetic field remains always normal to the boundary, we find 
\be
\dfrac{d{\cal E}_{\rm jet}}{dt}=-\displaystyle\iint \Pi\bm V \cdot d\bm a \,.
\label{eqdedt} \ee  

This has the simple meaning that the change in energy of the jet is due to the work done by the total pressure of its environment. 
On the perturbed boundary, we also have (see Section 3.2 in \cite{2023Univ....9..386V}) $\hat n\cdot {\bm V} 
=-i\omega \xi_L^{\hat \varpi} $, and thus, $\bm V \cdot d\bm a= -i\omega \xi_L^{\hat \varpi} da$ and 
$\Pi
=\Pi_0+\Pi_1 + \xi_L^{\hat \varpi} \dfrac{d \Pi_0}{d \varpi}$.
Using the functions $\xxx$ and $\yyy$, we can express 
$\bm V \cdot d\bm a= -i\omega \dfrac{\xxx}{\varpi} da \exp[i(m\phi+kz-\omega t)]$ and 
$\Pi
=\Pi_0+\yyy \exp[i(m\phi+kz-\omega t)]$.

Since we need to keep quadratic (nonlinear) terms in Equation~(\ref{eqdedt}), we should carefully replace the real parts of the functions and take the real part of the product. Taking the mean value of the result, the zeroth- and first-order terms disappear.\endnote{For two complex functions $A$ and $B$ that are proportional to $\exp(i\psi)$, the mean value of the product $\langle \Re A \, \Re B\rangle$ in an interval $\Re\psi\in (0,2\pi)$ is $\dfrac{1}{2}\Re[A^\ast B]$.} 
The mean value has a double meaning; over a length of the jet equal to multiples of the wavelength $2\pi/k$, or over a time period equal to multiples of $2\pi/\Re\omega$, provided that the growth time is much larger.   
The resulting expression for the mean value of the energy that is transferred from a length $\Delta z$ of the jet to its environment is
\be
\left\langle \dfrac{d{\cal E}_{\rm jet}}{dt} \right\rangle=\pi \, e^{2\Im \omega t}\displaystyle\iint \Re[i\omega \xxx\yyy^\ast] \dfrac{ da}{2\pi \varpi} 
= \pi\, e^{2\Im \omega t}  \Re[i\omega Y] \, |\yyy|^2\, \Delta z \,,
\label{eqdedt1}  
\ee
evaluated at $\varpi_j$.
This is related to the ``complementary energy'' derived in \cite{Goedbook2} in connection with the force operator and its non-self-adjointness.
The continuity of $Y$ and $\yyy$ ensures that the opposite energy per time is found when one integrates in the volume of the environment (because the area vector is opposite), and thus, the total energy remains constant. 

Finding the function $Y$ in the minimalist approach is sufficient to understand whether energy flows from the jet toward the environment, which is the case if $\Re Y \Im \omega+\Im Y \Re\omega>0$, or the opposite. 
If one needs the exact value, first, $\yyy$ needs to be found from Equation~(\ref{eqforxxxyyy}).
Obviously, Equation (\ref{eqdedt1}) also shows that the system is unstable if there exist at least one mode with positive $\Im\omega$.

The result can be directly generalized to any cylindrical distance, and the mean power transferred from the interior to the exterior (algebraically) at a radius $\varpi$ over a length $\Delta z$ is  
\be
\left\langle \dfrac{d{\cal E}(\varpi)}{dt} \right\rangle = \pi \, \Delta z \, e^{2\Im \omega t}  \Re[i\omega \xxx \yyy^\ast]\,.
\label{eqdedt2}  
\ee

We can also integrate this equation and find the mean energy contained between the axis and the cylindrical distance $\varpi$ over a length $\Delta z$:
\be
\langle {\cal E}(\varpi) \rangle = {\cal E}_0(\varpi)+\dfrac{\pi \, \Delta z}{2\Im \omega}\left(e^{2\Im \omega t} -1\right)  \Re[i\omega \xxx \yyy^\ast] \,,
\label{eqdedt3}  
\ee
where ${\cal E}_0(\varpi)$ is the energy of the unperturbed state and we assume $\langle {\cal E}(\varpi) \rangle|_{t=0} = {\cal E}_0(\varpi)$.

\section{An Example Case}\label{examplecase} 

In this section, the results of applying the minimalist approach to a model for relativistic magnetized jets are presented.
In particular, we chose to explore how relativity changes the results of the constant pitch magnetic field model considered in \cite{ALB00}. 
This is a simple model containing the basic characteristics of a jet and its environment. 
The goal is not only to discuss the physics involved---without, of course, being able to exhaust this interesting topic in this connection---but also to investigate the form of the eigenfunctions corresponding to different instability mechanisms using the new formalism.

In the unperturbed state, we assume that the jet extends up to a cylindrical distance $\varpi_j=1$, has a constant bulk velocity $V_0\hat z$ (Lorentz factor $\gamma_0$), has a constant rest density $\rho_{0a}$, has zero pressure $P_0=0$ (specific enthalpy $\xi_0=1$), and has a magnetic field 
$\bm B_0=B_a\dfrac{\hat z+(\gamma_0\varpi/\varpi_0)\hat\phi }{1+(\varpi/\varpi_0)^2}$ with constant $B_a$ (the field on the axis) and $\varpi_0$ (the pitch in the comoving frame).  
For the environment, we assume a static hydrodynamic medium with a rest density $\rho_{0e}=\eta \rho_{0a}$, pressure $P_{0e}=\dfrac{B_a^2/2}{1+(\varpi_j/\varpi_0)^2}$ from the pressure balance on the jet surface, and specific enthalpy  $\xi_{0e}=\dfrac{5\Theta_{0e}+\sqrt{9\Theta_{0e}^2+4}}{2}$, where $\Theta_{0e}=\dfrac{P_{0e}}{\rho_{0e}}$.

Relativity is included in the dynamics in three ways: by allowing the bulk, Alfv\'en, and sound velocities to be relativistic in general. 
Defining the Alfv\'en velocity on the axis $\Ua=\dfrac{B_a}{\sqrt{\rho_{0a}}}$ and the corresponding Mach number $M_{\rm A}=\dfrac{\gamma_0V_0}{\Ua}$,
the dimensionless parameters that fully define the unperturbed state are 
$\Ua$, $M_{\rm A}$, $\varpi_0$, and $\eta$. 

We use units defined in \cite{2023Univ....9..386V}, i.e., lengths in $\varpi_j$, wavelengths in $1/\varpi_j$, velocities in $c$, frequencies in $c/\varpi_j$, and the factor $\sqrt{4\pi}$ absorbed into the magnetic field.  
We additionally set $B_a=1$, and thus, we measure densities, energy densities, and pressures in units of $B_a^2$.

We assume in the following that $\Ua=1$, $M_{\rm A}=1$, $\varpi_0=0.33$, and $\eta=1$;
therefore, the bulk four velocity is $\gamma_0V_0=M_{\rm A}\Ua=1$, the Lorentz factor is $\gamma_0=\sqrt{1+\gamma_0^2V_0^2}=\sqrt{2}$, and the Alfv\'en three velocity on the axis is $v_{{\rm A}a}=\dfrac{\Ua}{\sqrt{1+\Ua^2}}=\dfrac{1}{\sqrt{2}}$.

\begin{figure}[H]
	\begin{minipage}{0.575\textwidth} 
		\includegraphics[width=.95\textwidth]{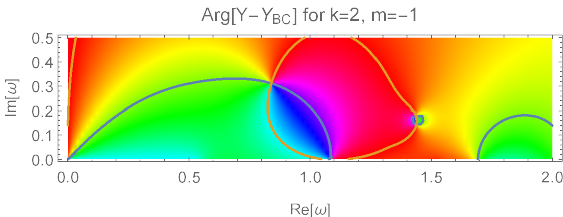}
		
		\vspace{3mm} 
		\includegraphics[width=.95\textwidth]{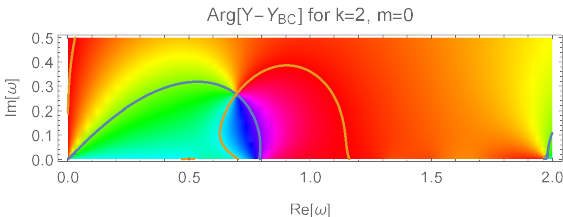}
		
		\vspace{3mm} 
		\includegraphics[width=.95\textwidth]{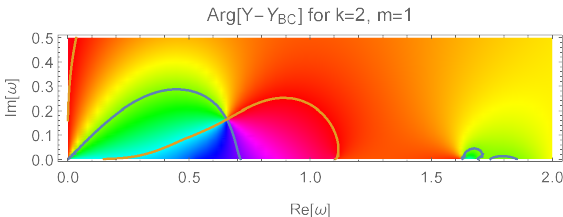}
	\end{minipage} 
	\hfill
	\begin{minipage}{0.4\textwidth} 
		\includegraphics[width=.95\textwidth]{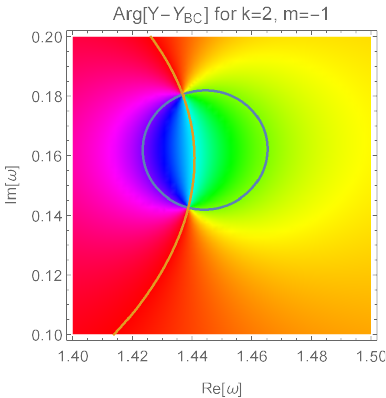}
		
		\vspace{17mm}
		\hfill 
		\includegraphics[width=.88\textwidth]{Figures/colorbar_arg}
		
		\vspace{10mm} $\ $
	\end{minipage}
	\caption{The Spectral Web. The parameters for each panel are shown at the top of the panel.
	The {\bf upper-right panel} is a zoomed region of the {\bf upper-left panel}.}  
	\label{figwebk2}
\end{figure}

Figure \ref{figwebk2} shows the Spectral Web for $k=2$ and $m=-1$, $0$, $1$ from top to bottom.

The top panel of Figure \ref{figwebk2} shows two eigenvalues, one at $\omega \approx 1.44+0.14i$ (shown more clearly in the upper right panel)
and a second at $\omega\approx 0.84+0.31i$. (It also shows two poles; we recall that ${\rm Arg}[Y-Y_{\rm BC}]$ increases as we move counterclockwise around the eigenvalues, while it decreases as we move counterclockwise around poles.)

The $m=-1$ has a sign opposite to the azimuthal magnetic field, making it possible to fulfill the resonant surface relations $\bm k_{\rm co}\cdot \bm B_{\rm co}=0 $ and $\omega_{\rm co}=0$ (for which ${\cal D}=0$). This is related to the current-driven instability (CDI). 
Thus, the eigenvalue $\omega \approx 1.44+0.14i$ corresponds to CDI since it is absent for $m=0$ and $m=+1$. 

The second eigenvalue corresponds to the Kelvin--Helmholtz instability (KHI). It is present for all $m$ (at a slightly shifted position) and is the so-called surface mode (SM), ordinary mode, or fundamental mode. 

\begin{figure}[H]
	\begin{minipage}{0.6\textwidth} 
		\includegraphics[width=.95\textwidth]{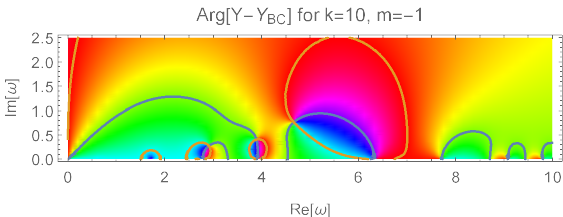}
		
		\vspace{3mm} 
		\includegraphics[width=.95\textwidth]{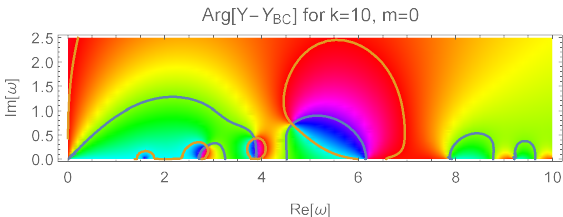}
		
		\vspace{3mm} 
		\includegraphics[width=.95\textwidth]{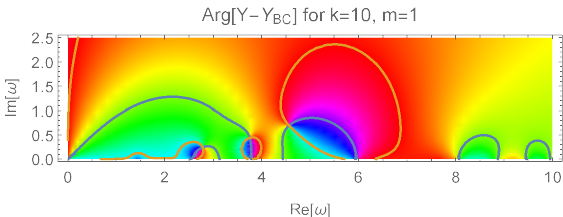}
	\end{minipage} 
	\hfill
	\begin{minipage}{0.4\textwidth} 
		\centering
		\includegraphics[width=.15\textwidth]{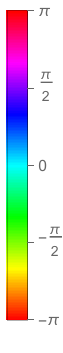} 
	\end{minipage}
	\caption{Similarly to Figure \ref{figwebk2} but for $k=10$. Compared with that figure, here we see the new body modes emerging at the left.}  
	\label{figwebk10}
\end{figure}

Figure \ref{figwebk10} shows the Spectral Web for $k=10$ and $m=-1$, $0$, $1$ from top to bottom.

We observe that the CDI is no longer present, which is something that is expected for sufficiently large $k$ beyond the value that satisfies the resonant surface relation. 

We also observe that new modes appear on the left of the SM (two modes in this case). 
These are connected to the Kelvin--Helmholtz instability and are called body modes (BMs) or reflection modes. 

The three panels are very similar, which is something that is expected for sufficiently large $k$ (compared with~$m$). 

\vspace{-8pt}

\begin{figure}[H]
\hspace{-10pt}	\includegraphics[width=.6\textwidth]{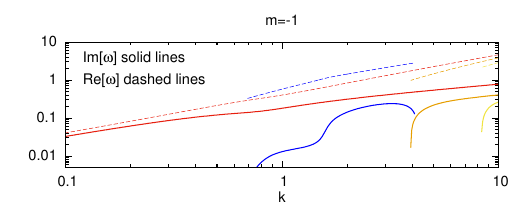}
	
	\vspace{0mm} 
\hspace{-10pt}	\includegraphics[width=.6\textwidth]{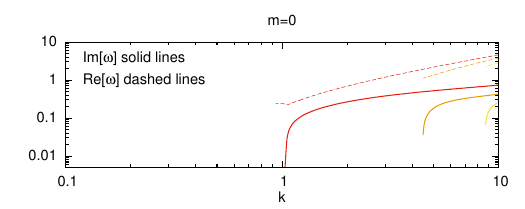}
	
	\vspace{0mm} 
\hspace{-10pt}	\includegraphics[width=.6\textwidth]{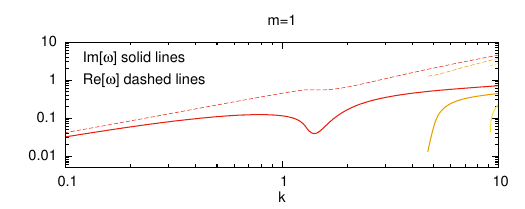} 
	\caption{The dispersion relation for $m=-1$ ({\bf top}), $m=0$ ({\bf middle}), and $m=1$ ({\bf bottom}).} 
	\label{figdispersion}
\end{figure}

Repeating the process for all $k$, we find the dispersion relation shown in Figure \ref{figdispersion} for the three values of $m$.

Regarding the CDI, which is the main focus in \cite{ALB00}, for the chosen parameters, relativity only slightly modify the results. 
The maximum growth rate is $\Im\omega\approx 0.24 $ at $k\approx 3.2$, where $\Re\omega=2.28$.
In the frame comoving with the jet, these translate to $\Re\omega_{\rm co}\approx 0$, $\Im\omega_{\rm co}=\gamma \Im\omega \approx 0.34 $, and $k_{{\rm co}z}\approx 2.26-0.24 i$.
The $\Im\omega_{\rm co}\dfrac{\varpi_j}{v_{{\rm A}a}}\approx 0.48$ is compared with the maximum growth rate shown in Figure 1 of \cite{ALB00}, i.e., $\Im\omega_{\rm co}\dfrac{\varpi_j}{v_{{\rm A}a}}\approx 0.4$ for $k_{{\rm co}z}\approx 2.3$.

As expected, the CDI is advected with the flow ($\Re\omega_{\rm co}\approx 0$) and the comoving wavevector along the motion is $k_{{\rm co}z}\approx \dfrac{k}{\gamma_0}-i\gamma_0V_0 \Im\omega$.
The resonant surface relations $\bm k_{\rm co}\cdot \bm B_{\rm co}=0 $ and $\omega_{\rm co}=0$ are shifted to $k=-\gamma_0\dfrac{m}{\varpi_0}\approx 4.24$, in agreement with the dispersion relation for CDI shown in the upper panel of Figure \ref{figdispersion} in blue (for $k$ larger than the one satisfying the resonant surface relation the mode is stable---this limiting $k$ depended on~$m$).

\begin{figure}[H]
		{\hspace{-5pt}\includegraphics[width=.4\textwidth]{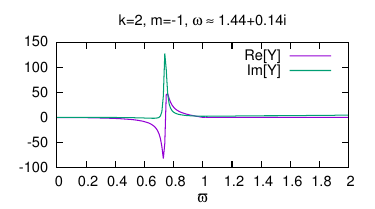}
		\hspace{2cm}	\includegraphics[width=.4\textwidth]{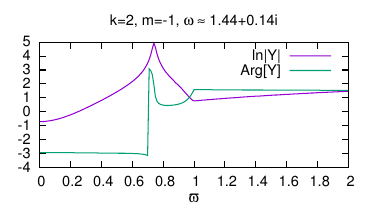}}
		
		\vspace{0mm} 
		{\hspace{-5pt}\includegraphics[width=.4\textwidth]{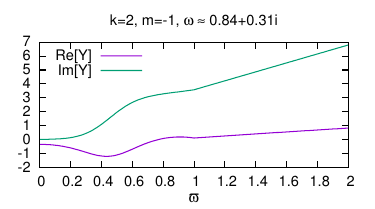}
		\hspace{2cm} \includegraphics[width=.4\textwidth]{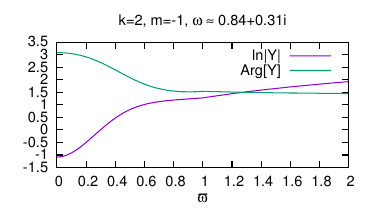}}
		
		\vspace{0mm} 
		{\hspace{-5pt}\includegraphics[width=.4\textwidth]{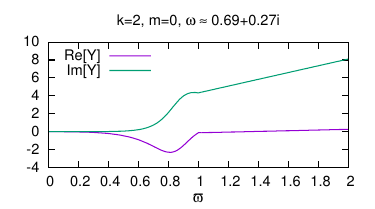}
		\hspace{2cm} \includegraphics[width=.4\textwidth]{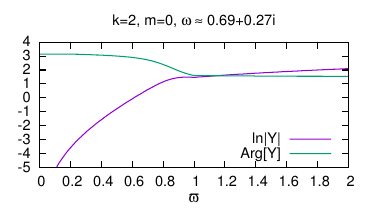}}
		
		\vspace{0mm} 
		{\hspace{-5pt}\includegraphics[width=.4\textwidth]{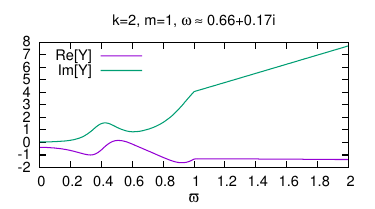} 
		\hspace{2cm} \includegraphics[width=.4\textwidth]{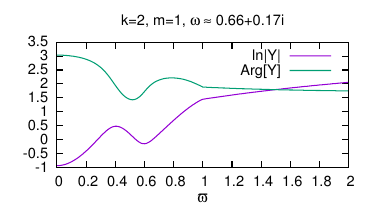}}
	\caption{The eigenfunctions for $k=2$. The parameters for each panel are shown at the top of the panel.} 
	\label{figeigenk2}
\end{figure}

The eigenfunctions for $k=2$ are shown in Figure \ref{figeigenk2}. 
The first row corresponds to the CDI (for $m=-1$). The following three rows correspond to the KHI modes for  the three values of $m$.

\begin{figure}[H]
	\hspace{-5pt}\includegraphics[width=.24\textwidth]{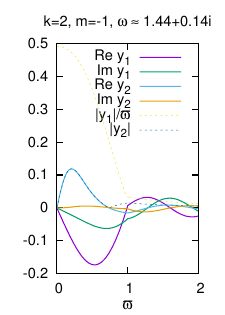}
	\hspace{-5pt}\includegraphics[width=.24\textwidth]{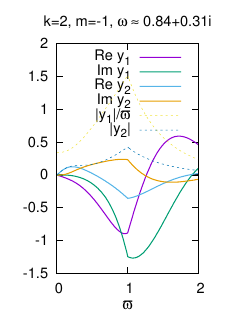}
	\hspace{-5pt}\includegraphics[width=.24\textwidth]{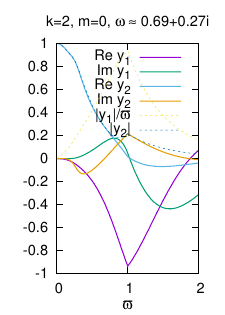} 
	\hspace{-5pt}\includegraphics[width=.24\textwidth]{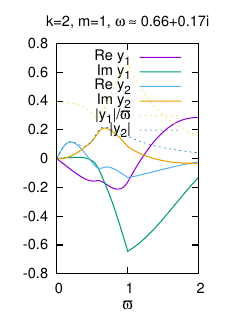}  
	\caption{The eigenfunctions $\xxx$ and $\yyy$ for $k=2$.
	}  
	\label{figeigenk2y1y2}
\end{figure}

The corresponding $\xxx$ and $\yyy$ are shown in Figure \ref{figeigenk2y1y2}. The absolute values of the amplitudes of the Lagrangian displacement $\dfrac{|\xxx|}{\varpi}$ and the total pressure $|\yyy|$ are also shown. These plots definitely provide additional information for the perturbation compared with what is shown in Figure \ref{figeigenk2} through $Y$. For $m=-1$, the CDI is mostly concentrated near the axis (and displaces the axis, as expected), while the KHI is near the interface. The modes for $m=0$ and $m=1$ show mixed behavior.
For $m=0$, the displacement is maximum at the interface, but the perturbation of the pressure is maximum on the axis. For $m=1$, we find displacement on the axis similarly to the CDI, but also on the interface. 

\vspace{-10pt}

\begin{figure}[H]
		{\hspace{-7pt}\includegraphics[width=.4\textwidth]{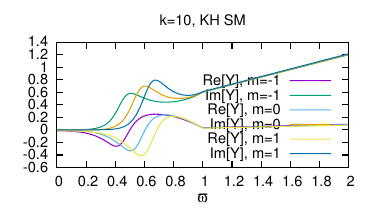}
		\hspace{2cm} \includegraphics[width=.4\textwidth]{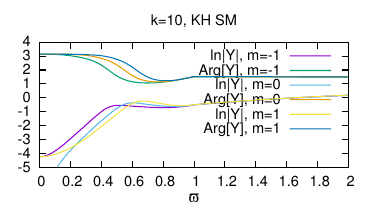}}
		
		\vspace{0mm} 
		{\hspace{-7pt}\includegraphics[width=.4\textwidth]{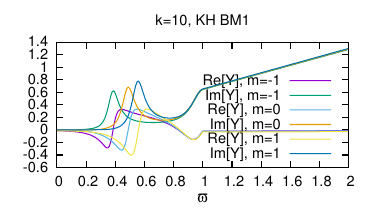}
		\hspace{2cm} \includegraphics[width=.4\textwidth]{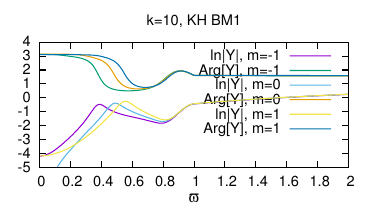}}
		
		\vspace{0mm} 
		{\hspace{-7pt}\includegraphics[width=.4\textwidth]{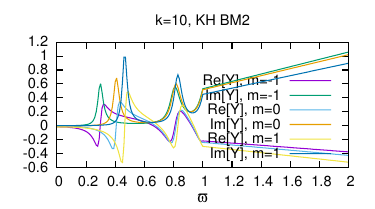}
		\hspace{2cm} \includegraphics[width=.4\textwidth]{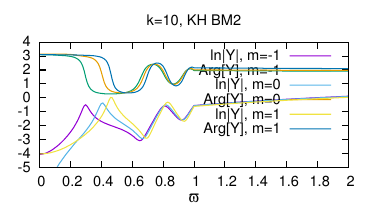}} 
	\caption{Similarly to Figure \ref{figeigenk2} but for $k=10$.
	}  
	\label{figeigenk10}
\end{figure}

Figure \ref{figeigenk10} shows the eigenfunctions for $k=10$. There are three eigenvalues for each value of $m$, which correspond, from larger to smaller $\Re\omega$, to SM, BM1, and BM2.
For large $k$, all $m$ give approximately the same result (eigenvalues and eigenfunctions). Counting seems to work well since more oscillations appear as we move to the next body mode. 
The general characteristics of the eigenfunctions, which are verified for even larger $k$ as well, are as follows: a region near the axis that is controlled by the boundary condition; then a region with oscillations, the number of which depends on the number of the mode; and then the region near the interface controlled by the boundary condition.

\begin{figure}[H]
	\hspace{-7pt}\includegraphics[width=.32\textwidth]{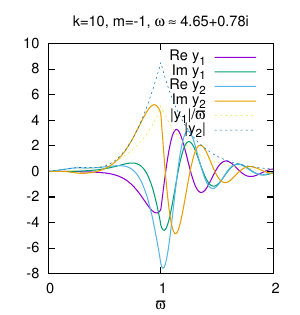}
	\hspace{-7pt}\includegraphics[width=.32\textwidth]{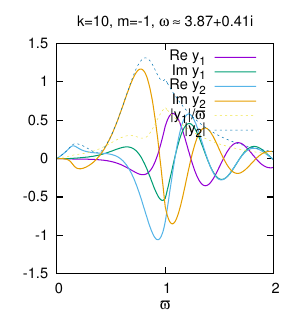} 
	\hspace{-7pt}\includegraphics[width=.32\textwidth]{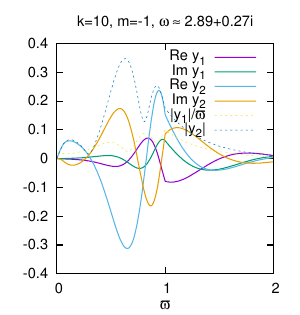}  
	\caption{The eigenfunctions $\xxx$ and $\yyy$ for $k=10$ and $m=-1$.
	}  
	\label{figeigenk10y1y2}
\end{figure}

The corresponding eigenfunctions $\xxx$ and $\yyy$ are shown in Figure \ref{figeigenk10y1y2} (the case $m=-1$ is shown, the others do not differ significantly due to the relatively large value of $k$). The numbers of oscillations are not as clear as in $Y$. They show, however, that the perturbations are more important near the interface. 

\vspace{-12pt}

\begin{figure}[H]
	\hspace{-0pt}\includegraphics[width=.5\textwidth]{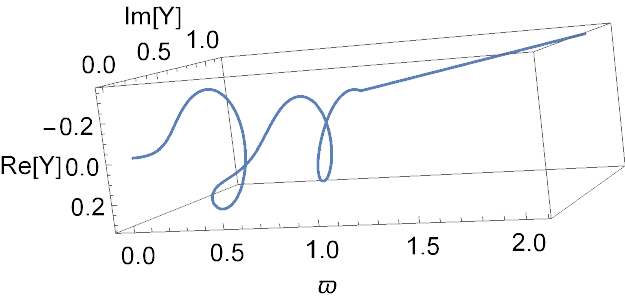} 
	\caption{
		The trajectory in the $Y$ plane for the BM2 for $k=10$ and $m=-1$.
		\label{figYplane}
	}  
\end{figure}

Clearly, the $Y$ on one hand and the $\xxx$ and $\yyy$ on the other give complementary information for the physics of the various unstable modes. 
However, the fact that the eigenvalues can be found in relation to $Y$ alone means that this function needs to be more carefully analyzed in the various cases. 
Its relation to the number of oscillations was already shown. 
Another way to see it is through the trajectory that the eigenfunction $Y$ follows in the complex $Y$ plane as $\varpi$ increases. For sufficiently large $k$, the number of windings equals the number of the mode.
An example is shown in Figure \ref{figYplane}. 

\section{Discussion}\label{secconclusions}

The minimalist approach presented in this paper offers a more economic way to find the eigenvalues of a linear stability problem by solving a single first-order differential equation for the complex function $Y$. 
Needless to say that although the presented examples are related to relativistic magnetohydrodynamic jets because this is perhaps the most challenging and unexplored area of stability, the minimalist approach can be applied in all kinds of fluid dynamic settings, simply by choosing the appropriate forms of the functions ${\cal F}_{ij}/{\cal D}$ and boundary conditions. For problems in cylindrical geometry, the formulas are given in \cite{2023Univ....9..386V} (these cover the nonrelativistic and the hydrodynamic limits). In other geometries, one can find these functions by linearizing the full equations. The method can also be applied to any other system that concerns the growth of perturbations in the linear~regime. 

A method to solve a complex boundary condition equation was developed by taking advantage of the fact that the function $Y$ is an analytic function of the complex eigenvalue; using an analogy with electrostatics; and, in essence, generalizing the ``Spectral Web'' method presented in \cite{Goedbook2}
by using only ${\rm Arg} [Y]$ and finding a way to distinguish roots and poles.
It is interesting that the minimalist process of finding the eigenvalues depends only on the phase difference between $\xxx$ and $\yyy$---which is the ${\rm Arg}[Y]$---and not the phases themselves nor the energies, although a connection is discussed in Section \ref{secenergy}.

The problem always leads to a complex equation $Y-Y_{\rm BC}=0$, but there are many variations of this equation. For example, in the case presented in \cite{2023Univ....9..386V}, the mathematical expression of the equation is different if we apply it to the wall interface $\varpi=1$ (having integrated the equation from the axis to this interface), or in the interface $\varpi=0.5$ (having integrated the equation from the axis to this interface and from the wall to this interface separately). Although the poles are different, the roots are always the same and equal to the eigenvalues of the problem. 

To obtain the full picture of an instability, we of course also need $\xxx$ or $\yyy$ (in addition to their ratio $Y$). For example, for the energy consideration, we saw that by knowing $Y$, we find the direction of the energy flux but not the magnitude and its radial dependence. Also, the eigenfunction $Y$ cannot say whether, e.g., there is a common dropping factor in both $\xxx$ and $\yyy$ since it is their ratio and the factor cancels out. We provide such an example in Appendix \ref{appendixapprox}. In addition, some connection with $\xxx$ and $\yyy$ is hidden inside the boundary conditions on the axis and at large distances. When we derived these conditions, we used information for $\xxx$ and $\yyy$ (to not diverge on the axis and to vanish at infinity, representing outgoing waves). However, this was done once, and since we know the boundary conditions, we are set to use the minimalist approach and the rest of the process uses only $Y$. 
This is advantageous because it more efficiently solves the most difficult part of the problem, which is to find the eigenvalues. Once we know an eigenvalue, it is trivial to return to the full system and calculate the perturbations of all quantities to obtain an overall picture of the eigenstate. 

The fact that $Y$ does not contain common factors of $\xxx$ and $\yyy$ also has its advantages. By dropping the common factors that affect the amplitudes, we concentrate on the phase of the oscillations, and this is why $Y$ is apparently the function more closely related to the characterization of the body modes through the number of their oscillations. The oscillations are shown more clearly through $Y$, not only because the dropping factors are missing but also because the wavelength is smaller compared with the wavelength of $\xxx$ and $\yyy$, as shown in the example in Appendix \ref{appendixapprox}.

Note also that the oscillatory behavior is expected in general for large $k$ and for Kelvin--Helmholtz instability modes, but not necessarily in current-driven instability modes.
There are interesting topics to be explored further, e.g., the connection between the solutions for $Y$ and the function $\tilde\kappa$, and the phase difference between $\xxx$ and $\yyy$ in relation to the phase of each function separately, but we leave these for future studies.
Although we already commented about the physics of the perturbations in various aspects, the goal of this paper was mainly to present the minimalist approach and the way to find the eigenvalues using it, which can be seen as the starting point for the rest. 

We recall that the functions $\xxx$ and $\yyy$ are solutions of the linear problem, and thus, they can be freely multiplied with a complex constant (the same for both). Thus, we can freely multiply their amplitudes with a constant number and shift their phases with a constant angle. Their ratio, namely, the eigenfunction $Y$, is uniquely defined.
\\ In case we need to recover the units, $\xxx$ has units of length squared, and thus, we should multiply it by $\varpi_j^2$, and $\yyy$ has units of pressure, and thus, we should multiply it by $B_a^2$ (which includes a factor of $4\pi$, and thus, it is twice the magnetic pressure on the axis).

\vspace{6pt} 




\funding{This research received no external funding.}

\dataavailability{This research is analytical; no new data were generated or analyzed. If needed, more details on the study and the numerical results will be shared upon reasonable request to the author.} 


\conflictsofinterest{The author declares no conflicts of interest.} 

\appendixtitles{yes} 
\appendixstart
\appendix


\section{Typical Behavior of \boldmath{$Y$}}\label{appendixapprox}

Noting the dominant dependences in ${\cal F}_{ij}/{\cal D}$, as given by Equations (54)--(57) in \cite{2023Univ....9..386V}, and assuming
$ \dfrac{{\cal F}_{11}}{\cal D}=\dfrac{A}{\varpi}$, $\dfrac{{\cal F}_{12}}{\cal D}=B\varpi $, $ \dfrac{{\cal F}_{21}}{\cal D}=-\dfrac{\tilde\kappa^2}{B\varpi}-\dfrac{C}{B\varpi^3}$, and $ \dfrac{{\cal F}_{22}}{\cal D}=-\dfrac{A}{\varpi}$
with constant $A$, $B$, $C$, and $\tilde\kappa$, 
we obtain the analytical solution 
$Y
=\dfrac{-\varpi{\cal F}_{12}/{\cal D}}{A+1-\nu+\tilde\kappa\varpi\dfrac{C_1J_{\nu-1}(\tilde\kappa\varpi)+C_2Y_{\nu-1}(\tilde\kappa\varpi)}{C_1J_\nu(\tilde\kappa\varpi)+C_2Y_\nu(\tilde\kappa\varpi)}}$ with $\nu=\sqrt{(A+1)^2-C}$.
The corresponding $\xxx$ is $\xxx=C_1\varpi J_\nu(\tilde\kappa\varpi)+C_2\varpi Y_\nu(\tilde\kappa\varpi)$.
\\ For large $|\tilde\kappa|\varpi$,
we can further simplify the expressions $ \dfrac{{\cal F}_{11}}{\cal D}=0$, $\dfrac{{\cal F}_{12}}{\cal D}=B\varpi $, $ \dfrac{{\cal F}_{21}}{\cal D}=-\dfrac{\tilde\kappa^2}{B\varpi} $, and $ \dfrac{{\cal F}_{22}}{\cal D}=0$, 
and obtain the solution
$Y=-\dfrac{{\cal F}_{12}}{\tilde\kappa{\cal D}}\dfrac{C_1J_1(\tilde\kappa\varpi)+C_2Y_1(\tilde\kappa\varpi)}{C_1J_0(\tilde\kappa\varpi)+C_2Y_0(\tilde\kappa\varpi)}
\approx \dfrac{{\cal F}_{12}}{\tilde\kappa{\cal D}}\cot(\tilde\kappa\varpi+\phi_0) $,
with $\phi_0=\dfrac{\pi}{4}+\arctan\dfrac{C_2}{C_1}$ using the Bessel asymptotics,
essentially corresponding to a harmonic oscillator with a complex frequency.
For $\xxx$ and $\yyy$, we obtain $\xxx=C_1\varpi J_1(\tilde\kappa\varpi)+C_2\varpi Y_1(\tilde\kappa\varpi)
\approx -\sqrt{\dfrac{2(C_1^2+C_2^2)}{\pi\tilde\kappa}} \sqrt{\varpi}\cos(\tilde\kappa\varpi+\phi_0)$
and $\yyy=\dfrac{\xxx}{Y}=-\sqrt{\dfrac{2(C_1^2+C_2^2)}{\pi\tilde\kappa}}  \dfrac{\tilde\kappa}{B\sqrt{\varpi}}\sin(\tilde\kappa\varpi+\phi_0)$.
 
The function $Y$ is proportional to $\dfrac{\cot(\tilde\kappa\varpi+\phi_0)}{\tilde\kappa}$, which can be written as
	\begin{eqnarray}
\dfrac{\cot(\tilde\kappa\varpi+\phi_0)}{\tilde\kappa}
=\dfrac{1}{|\tilde\kappa|^2}\dfrac{\Re \tilde\kappa\sin (2\Re \tilde\kappa\varpi+2\Re \phi_0)-\Im\tilde\kappa \sinh (2\Im \tilde\kappa\varpi+2\Im \phi_0) }{\cosh(2\Im \tilde\kappa\varpi+2\Im \phi_0)-\cos(2\Re \tilde\kappa\varpi+2\Re \phi_0)}
\nonumber \\ 
-i\dfrac{\Re\tilde\kappa \, \Im\tilde\kappa}{|\tilde\kappa|^2}\dfrac{
	\dfrac{\sinh (2\Im \tilde\kappa\varpi+2\Im \phi_0)}{\Im\tilde\kappa}+\dfrac{\sin (2\Re \tilde\kappa\varpi+2\Re \phi_0)}{\Re \tilde\kappa}}{\cosh(2\Im \tilde\kappa\varpi+2\Im \phi_0)-\cos(2\Re \tilde\kappa\varpi+2\Re \phi_0)}
\,. 
	\end{eqnarray}

We can see that the $\Re \tilde\kappa\varpi$ dependence creates oscillations in $Y$ with wavelength $\pi/\Re\tilde\kappa$ and the $\Im \tilde\kappa\varpi$ dependence affects their amplitude. 
{The imaginary part of $\dfrac{\cot(\tilde\kappa\varpi+\phi_0)}{\tilde\kappa}$ has a sign that is controlled by the sign of $\Im\left[\tilde\kappa^2\right]$,
and its absolute value peaks whenever $2 \tilde\kappa\varpi+2 \phi_0$ approaches even multiples of $\pi$.}

The functions $\xxx$ and $\yyy$ have an additional common factor $1/\sqrt{\varpi}$ in their amplitudes and they oscillate with wavelength $2\pi/\Re\tilde\kappa$ and some difference in their phases.

\begin{figure}
	\includegraphics[width=.29\textwidth]{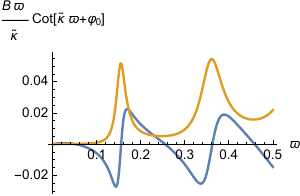}\includegraphics[width=.29\textwidth]{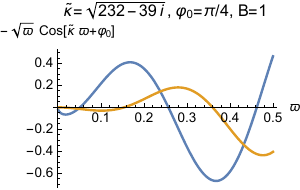}\includegraphics[width=.41\textwidth]{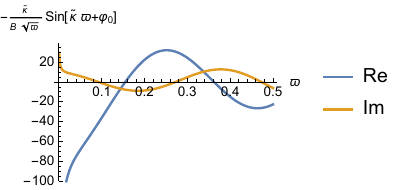}
	\caption{A solution typical for cases with $|\tilde\kappa|\varpi$ larger than unity. 
		Approximations are shown for $Y$ ({\bf left}), $\xxx$ ({\bf middle}), and $\yyy$ ({\bf right}).
		\label{figfitexampleY}
	}  
\end{figure}

An example can be seen in Figure \ref{figfitexampleY}.  
This behavior turns out to be a very good approximation of exact solutions with $|\tilde\kappa|\varpi$ larger than unity. Actually, the parameters for the solution in Figure \ref{figfitexampleY} were chosen to fit an exact solution of the problem from \cite{2023Univ....9..386V}, as shown in the bottom-right panel of Figure \ref{figwebpapertheory}
(the bottom-left panel of Figure \ref{figfitexampleY} and the 
bottom-right panel of Figure \ref{figwebpapertheory} are practically indistinguishable in the region $\varpi<0.5$). 

Of course in the general case, $\tilde\kappa$ is a function of $\varpi$ and the wavelength of the oscillations is variable. In this case, the basic characteristics of the solution can be understood by replacing the argument of the $\tan$ with $\displaystyle\int\tilde\kappa\, d\varpi$.

If $|\tilde\kappa|\varpi$ is smaller than unity, the above approximations are not valid. This is the case at least in a small region near the axis, in which we know the behavior of the solution from the analysis of the boundary condition in Appendix \ref{secboundaxis}.

\section{Details on the Boundary Conditions} 

\subsection{Boundary Conditions on the Axis}\label{secboundaxis}
\begin{itemize}
	\item For $m\neq 0$, the boundary condition at the symmetry axis is $Y(\varpi=0)=\dfrac{\lambda_1}{\lambda_2}$; see Equation (34) of \cite{2023Univ....9..386V}.
	\\ 
	This is enough information for practical purposes; nevertheless, it is worth examining the behavior of $Y$ near the axis. 
	\\ For $m\neq 0$, all limits $d_{ij}=\displaystyle\lim_{\varpi\to 0}\dfrac{\varpi {\cal F}_{ij}}{\cal D}$ are constants (given in Appendix B of \cite{2023Univ....9..386V}), and the relations $d_{22}=-d_{11} $ and 
	$d_{11}^2+d_{12}d_{21} =m^2$ hold; thus, the equation for $Y$ becomes $\dfrac{\varpi}{|m|}\dfrac{d F}{d\varpi}=F^2-1$, with $F=\dfrac{d_{21}Y-d_{11}}{|m|} $ or $\dfrac{d_{12}/Y+d_{11}}{|m|}$.
	The acceptable solution is
	$F=-1\Leftrightarrow \dfrac{\varpi{\cal F}_{21}}{\cal D}Y-\dfrac{\varpi{\cal F}_{11}}{\cal D}=-|m|\Leftrightarrow \dfrac{\varpi{\cal F}_{12}}{{\cal D}Y}+\dfrac{\varpi{\cal F}_{11}}{\cal D}=-|m| $ because, according to Equation~(\ref{eqforxxxyyy}), only this corresponds to finite $\xxx$ and $\yyy$ on the axis. 
	There is also the unacceptable solution $F=1\Leftrightarrow\dfrac{\varpi{\cal F}_{12}}{{\cal D}Y}+\dfrac{\varpi{\cal F}_{11}}{\cal D}=+|m|$ corresponding to $\xxx\propto \varpi^{-|m|}$ and $\yyy\propto \varpi^{-|m|}$.
	The boundary can be seen as a regularity condition to choose the acceptable solution $Y=-\dfrac{d_{12}}{d_{11}+|m|}=\dfrac{d_{11}-|m|}{d_{21}}$
	(and not the unacceptable $Y=\dfrac{d_{12}}{|m|-d_{11}}=\dfrac{|m|+d_{11}}{d_{21}}$).
	\\ Actually, the general solution of $\dfrac{\varpi}{|m|}\dfrac{d F}{d\varpi}=F^2-1$ is $F=\dfrac{1+(\varpi/\varpi_0)^{2|m|}}{1-(\varpi/\varpi_0)^{2|m|}}
	=\coth x$ and $x=-|m|\ln \dfrac{\varpi}{\varpi_0}$, and thus, the acceptable solution $F=-1$ corresponds to $\varpi_0=0$.
	\\ Concluding, for $m\neq 0$, the boundary condition at the symmetry axis is the one given in Equation (\ref{eqbcaxismneq0}) of the main text.
	\item For $m= 0$, the boundary condition at the symmetry axis is $Y(\varpi=0) =-\dfrac{b_{12}}{2}\varpi^2 $; see Equation (35) of \cite{2023Univ....9..386V}.
	For practical purposes, it is enough to assume $Y(\varpi=0)=0$. 
	\\ In more detail, for $m=0$, the constant limits are 
	$b_{11}=\displaystyle\lim_{\varpi\to 0}\dfrac{{\cal F}_{11}}{\varpi {\cal D}}$,
	$b_{12}=\displaystyle\lim_{\varpi\to 0}\dfrac{{\cal F}_{12}}{\varpi {\cal D}}$,
	$b_{21}=\displaystyle\lim_{\varpi\to 0}\dfrac{\varpi {\cal F}_{21}}{{\cal D}}$, and
	$b_{22}=\displaystyle\lim_{\varpi\to 0}\dfrac{{\cal F}_{22}}{\varpi {\cal D}}$
	(see Appendix B of \cite{2023Univ....9..386V}), and
	the equation near the axis becomes 
	$ \dfrac{dY}{d\varpi}=b_{21}\dfrac{Y^2}{\varpi}+(b_{22}-b_{11})\varpi Y-b_{12}\varpi $. This has the acceptable solution $Y=-\dfrac{b_{12}\varpi^2}{2}$ corresponding to $\yyy=$ constant and $\xxx=-\dfrac{b_{12}\varpi^2}{2}\yyy$, but also the unacceptable $Y=-\dfrac{1}{b_{21}\ln\varpi}$ corresponding to $\xxx=$ constant and $\yyy=-\xxx b_{21}\ln\varpi$.
	The boundary condition can be seen as a regularity to choose the acceptable solution $Y=-\dfrac{b_{12}\varpi^2}{2}$.
	\\ Actually, we can find the general solution by noting that the term with $\varpi Y$ is always negligible compared with the largest between the $Y^2/\varpi$ and $\varpi$, and thus, the differential equation can be approximated as $ b_{21}\varpi\dfrac{dY}{d\varpi}=b_{21}^2Y^2+\lambda^2\varpi^2$, with $\lambda^2=-b_{12}b_{21}$. 
	The exact solution is $b_{21}Y=\lambda \varpi \dfrac{J_1(\lambda\varpi)+CY_1(\lambda\varpi)}{J_0(\lambda\varpi)+CY_0(\lambda\varpi)}$.
	The acceptable solution corresponds to $C=0$ and the unacceptable solution to $C=\infty$. 
	\\ Concluding, for $m= 0$, the boundary condition at the symmetry axis is the one given in Equation (\ref{eqbcaxismeq0}) of the main text.
\end{itemize}

\subsection{Boundary Conditions at Infinity}\label{secboundinfinity}

At large distances from the axis, assuming zero velocity and homogeneous medium with zero $B_{0\phi}$, 
we have the case of Section 5.1 of \cite{2023Univ....9..386V}.
The solution is 
\be 
Y=\dfrac{\cal D}{\varpi{\cal F}_{21}}\left[\dfrac{\lambda\varpi \, H_{|m|+1}^{(1)}(\lambda\varpi)}{H_{|m|}^{(1)}(\lambda\varpi)}-|m|\right]
\,, \ee
with constant $\dfrac{{\cal D}}{\varpi{\cal F}_{21}}$ and $\lambda$. It is acceptable if $\Im\lambda\ge 0$ (such that the amplitude of $\yyy$ does not diverge for $\varpi\to\infty$) and 
$\Re\lambda$ has the sign of $\Re\omega$, corresponding to outgoing waves.
Note that asymptotically, the ratio of the Hankel functions in the above equations approaches $-i$; thus, for $|\lambda|\varpi\gg 1$, we have 
\be 
Y\approx \dfrac{\cal D}{\varpi{\cal F}_{21}}\left(-i \lambda\varpi-|m|\right)
\,. \ee

More generally, whatever is the unperturbed state of the jet environment, the perturbation should vanish at infinity, and if it is oscillating, it should correspond to waves propagating toward larger $\varpi$. 

\section{Integration through Infinities of \boldmath{$Y$}}\label{appendixpole}
	
	Infinities of $Y$ are rare since they correspond to the vanishing of $\Re\yyy$ and $\Im\yyy$ simultaneously, but in general, it is possible at any distance $\varpi$ for particular values of $\omega$. They correspond to poles in the $\omega$ plane, and require fine tuning to find them, similar to the process we follow at the distance $\varpi_j$ for the function $Y-Y_{\rm BC}$.
	The needed perfect accuracy means that $Y$ never becomes infinity and the numerical integration passes through such points without a problem.  
	
\begin{figure}[H]
	{\includegraphics[width=.82\textwidth]{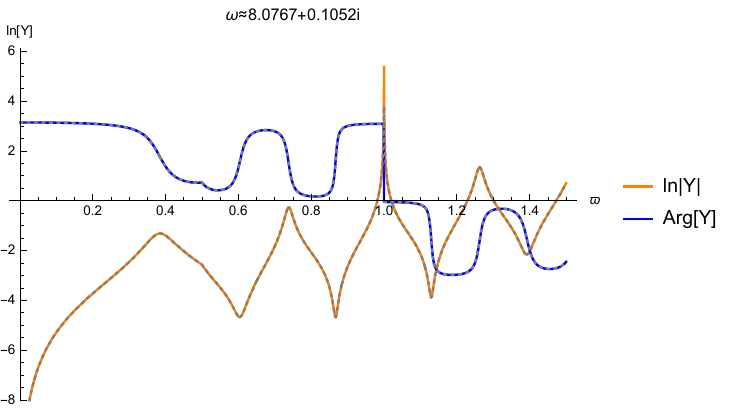}}
	\caption{Integration through a pole at the distance $\varpi=1$. 
	The two parts of the function $Y$ are shown: the ${\rm ln}|Y|$, which becomes infinity at the pole, and the ${\rm Arg}[Y]$, whose $\pi$ jump corresponds to the change in sign of $\yyy$.
	The solid (orange and blue) lines correspond to the integration of Equation~(\ref{eqforY}) and the dotted gray lines (that are practically on top of the solid lines) to the integration of the linear system~(\ref{systemodes}). 
	\label{figpole} }  
	\end{figure}

	Figure \ref{figpole} shows an example. It corresponds to the case of Figure \ref{figwebpapertheory}. As seen in the top-left panel of that figure, by integrating from the axis, we find a pole at $\varpi=1$ for $\omega\approx 8.1+0.1i$. To explore how the numerical integration behaved around the pole, we continued the integration for $\varpi>1$.
	The resulting $Y$ is shown in Figure \ref{figpole}.
	Even when we fine tuned the value of $\omega$, approaching as much as possible to the value corresponding to the pole, the integration continued without a problem. The $\ln |Y|$ became large, but not infinity, and the ${\rm Arg}[Y]$ approached a step function, but its variation was smooth, as can be seen by zooming close to the point $\varpi=1$. 
	Comparing the result of the integration with the integration of the linear system (\ref{systemodes})---or with the analytical expressions that exist for this particular model, see Ref. \cite{2023Univ....9..386V}---we found indistinguishable results (shown in the Figure \ref{figpole} as dotted gray lines). 
	
	This example shows that even if the integration encounters a pole at some distance, it is capable of resolving the differential equation for $Y$ around this point and gives the correct solution at larger distances.  
	
\begin{adjustwidth}{-\extralength}{0cm}
\printendnotes[custom] 

\reftitle{References} 




\PublishersNote{}
\end{adjustwidth}
\end{document}